\documentclass[12pt]{article} 
\usepackage{amssymb} 
\def\M{{\mathcal M}} 
\def\X{{{\mathbb R}^+}}
\def\Y{{\mathcal Y}}

\def\PHI{{\widehat {\Phi}}}

\def\Varphi{{\widehat\varphi}}

\newcommand{\be}{\begin{equation}} 
\newcommand{\ee}{\end{equation}}
\newcommand{\ea}{\end{eqnarray}} 
\newcommand{\pa}{\partial} 
 
\def\vrul{\rule[20pt]{0pt}{0pt}}

\def\bea{\begin{eqnarray}} 
\def\eea{\end{eqnarray}} 
\def\le{\left} 
\def\ri{\right}

\def\l{\lambda}

\def\R{{\mathbb R}}

\def\bes{$$} 
\def\ees{$$} 
\def\beas{\begin{eqnarray*}} 
\def\eeas{\end{eqnarray*}}

\newlength{\Mylen} 
 
\begin{document} 

\title{Correspondence between Minkowski and  de Sitter Quantum Field Theory}
\author{Marco Bertola$^{\rm a}$,
 Vittorio Gorini$^{\rm b}$, Ugo Moschella$^{\rm b,c}$, Richard Schaeffer$^{\rm
c}$}
\maketitle
\centerline {$^{\rm a}$ {SISSA, v. Beirut 2--4, 34014 Trieste}}\vskip 3pt
\centerline {$^{\rm b}$ Dipartimento di Scienze Matematiche Fisiche e
Chimiche,}
\centerline{ Via 
Lucini 3, 22100 Como and INFN sez. di Milano, Italy}\vskip 3pt
\centerline {$^{\rm c}$ {Service de Physique Th\'eorique, C.E. Saclay,
91191 Gif-sur-Yvette, France}}\vskip 3pt\vskip 8pt
\hrule
\begin{abstract}
In this letter we show that  the ``preferred'' 
Klein--Gordon Quantum Field Theories (QFT's)
 on a $d$-dimensional de Sitter spacetime 
can be obtained from a Klein--Gordon QFT
on a ($d+1$)-dimensional ``ambient'' Minkowski spacetime satisfying
the spectral condition 
and, conversely, that a  Klein--Gordon QFT 
on a ($d+1$)-dimensional ``ambient'' Minkowski spacetime
satisfying
the spectral condition can be obtained as superposition of $d$-dimensional
de Sitter Klein--Gordon fields in the preferred vacuum.
These results establish a correspondence between QFT's living on manifolds
having different dimensions. The method exposed here can be applied to study
other situations and notably QFT on Anti de Sitter spacetime. 
\end{abstract}
\hrule

\vskip40pt 
The study of the relations  
between Quantum Field Theories (QFT's)
in different dimensions has  
come recently to the general  
attention. Some of the most 
interesting and intriguing developments 
of QFT and string theory,
like Maldacena's ADS/CFT conjecture \cite{maldacena,witten}
and t'Hooft's \cite{thooft}
and Susskind's \cite{susskind}
holographic principle,  
seem to indicate that   
relations of this kind are going to play a fundamental 
role in understanding
QFT and string theory.

In this letter we point out a relation that exists between 
Minkowski QFT and de Sitter QFT in one dimension less.
We  show that the  ``preferred''  \cite{gibbonshawking,bunchdavies}
de Sitter Klein--Gordon field of squared mass $\lambda$ 
arises by averaging in a well-defined sense 
an ordinary Klein--Gordon field  of mass $M$ living in the Minkowski 
ambient spacetime and, vice versa, the Klein--Gordon field in the ambient Minkowski spacetime can be obtained by superposing 
fields in the lower dimensional de Sitter manifold.

The idea that QFT's on the de Sitter manifold can be obtained 
by restriction from the ambient spacetime is of course
not new, but it has been of little use in the standard 
coordinate approach to de Sitter field theories. 

Recently,  however, it has been shown 
\cite{brosgazeaumoschella,brosmoschella} that 
the well-known  \cite{gibbonshawking}
thermal properties 
of the de Sitter Klein--Gordon fields  in the ``preferred'' vacuum, 
are linked to certain analyticity properties 
of the correlation functions; 
these properties are precisely obtained by  
restriction to the de Sitter manifold 
of the  analyticity properties
of the (general) correlation functions in the ambient spacetime,   
which hold when the corresponding QFT satisfies
the energy--momentum spectral condition \cite{sw}.

This idea has then been  pushed further and it has become possible to show
that the thermal interpretation 
can be established also 
for interacting de Sitter field theories
\cite{brosepsteinmoschella}.

In this letter we take one step more
by showing how a Klein--Gordon Minkowski field in the Wightman vacuum
gives rise to  a Klein--Gordon  field on the de Sitter spacetime  
in the ``preferred'' thermal vacuum, giving a further argument in favour
of the adjective ``preferred''. Indeed, as it is well known,
there are in general infinitely many inequivalent vacua corresponding
to a certain QFT and one needs criteria to select the physically meaningful  ones.
This is true already for Minkowski QFT  
but, in this case, one has strong
physical criteria to select among the vacua. 
The situation is more difficult when considering
QFT's on a curved background, but several criteria have been established
also in this case (many of them however work only for linear field theories).
The ``preferred'' vacuum for de Sitter Klein--Gordon fields  
has been shown to satisfy many of such criteria, 
such as the  Hadamard condition  \cite{wald}. 
Here we prove that, actually, 
the ``preferred'' Klein--Gordon de Sitter QFT's
can be directly obtained
by  any massive or massless 
Wightman Klein--Gordon QFT in the ambient spacetime.
There is however a restriction on the mass of the fields that can be
obtained this way. 
In particular when working with a $d$-dimensional 
de Sitter hyperboloid of unit radius 
we can construct  
de Sitter Klein--Gordon fields whose mass is greater or equal to $(d-1)/2$.

The existence of such a link  
gives also a quantitative support to  the idea that a thermal effect on 
a curved manifold can be looked at as an Unruh effect
in a higher (flat) dimensional spacetime \cite{bertola,deser}.

Let therefore $\M=\{X \in {\mathbb M}^{d+1}:\ \eta_{\mu\nu}X^\mu X^\nu<0 \}$ 
be the manifold of events which are spacelike w.r.t. a chosen event (which
is taken as the origin of a frame) 
of a  $(d + 1)$-dimensional Minkowski space-time 
${\mathbb M}^{d+1}$. 
Coordinates of ${\mathbb M}^{d+1}$ are denoted by $\{X^\mu\}$, $\mu=0,\dots, d$.
$\M$ is foliated by  a family
of 
$d$-dimensional de Sitter spacetimes identified with 
the hyperboloids   
\begin{equation}
\Y_R = \{\eta_{\mu\nu} X^\mu X^\nu  = (X^0)^2-(\vec X)^2 = -R^2\ \}.
\end{equation} 
As a topological manifold, $\M= \X \times \Y$, where 
$\X$ is the  positive real half-line with coordinate 
$R$ and $\Y = \Y_1$ is the $d$-dimensional de Sitter spacetime 
with radius $R=1$. 
Points of $\Y$ are denoted by $y$ (i.e. $y^2 = -1$).

We can therefore assign to an event of $\M$ coordinates $(R,y)$ so that 
$X=Ry$ and $y^2=-1$. 
The Minkowskian metric of $\M$ can consequently  be rewritten as follows:
$$ 
ds^2=-dR^2+R^2 ds_{\Y}^2\ , 
$$ 
where $ds_{\Y}^2$ is the de Sitter  metric of $\Y$  
obtained as restriction of the Minkowski metric of the ambient space.

$\M$ is a globally hyperbolic  manifold where 
quantum field theory can be formulated \cite{wald}.
Let us therefore consider a 
{canonical} quantum field $\PHI$ on $\M$ satisfying 
 the Klein--Gordon equation
$(\square+M^2)\PHI(X)=0$ (in the following we will consider
the massive case $M>0$; the massless case can be obtained by a limit procedure
but can also be studied directly, with a considerable simplification of the formulae), and let 
us also  consider the corresponding equation for the modes $\Phi(X)$.
By separating the variables as in the metrics we write 
$\Phi(X)=\theta_\lambda(R)\varphi_\lambda(y)$
and we are led to  the following  equations:
\begin{eqnarray} 
&& \left(\square_\Y + \lambda\right)\varphi(y) = 0 \label{kgds}
\\
&& 
R^2\left(\pa_R^2+\frac d R \pa_R-M^2\right)\theta_\l (R)=-\lambda 
\theta_\l (R)\ .\label{desittertransverse} 
\end{eqnarray}
Let us consider in particular the equation (\ref{desittertransverse})
for the radial modes $\theta_\l$.
The operator appearing at the L.H.S.    
is  self-adjoint (on a suitable domain)
w.r.t. the following Hilbert product:  
\be
(\theta,\eta)=\int_{\X}\overline\theta (R)\eta 
(R)R^{d-2} dR
\ . 
\ee 
By means of the  transformation 
$ 
\theta(R)=R^{\frac{1-d}2}f(R)\label {lommel} 
$ 
and the  rescaling 
$\rho=MR$, which together are particular instances of the so called 
`Lommel's transformation', 
Eq. (\ref{desittertransverse}) 
is  turned  into the modified Bessel's equation.
By further introducing  the variable $x=\log \rho=$,  
we finally
obtain the following equation  
(the prime means derivative w.r.t. $x$):
\be
-f_\l''+(e^{2x}-\nu^2)f_\l=0 \label{expowall} \qquad \hbox{with}\qquad 
\nu=\nu(\lambda) = \sqrt{\lambda-\frac{(d-1)^2}4}\label{nu}.
\ee
We have thus obtained the Schr\"odinger 
problem for a particle in a one-dimensional potential $e^{2x}$ with eigenvalue $\nu^2$; 
this problem is to be studied 
in the standard  Hilbert space $L^2(\R)$. 
The spectrum   is nondegenerate and coincides with 
the positive real line;  this implies that $4\lambda>(d-1)^2$. 

The solutions 
which have the correct asymptotic behaviour at $x=\infty$ 
are  the modified Bessel functions  
$K_{i\nu}(e^x)$ \cite{bateman}. These modes are real, and 
their  asymptotic behaviour  near $x=-\infty$ 
is the following: 
\be 
K_{i\nu}(e^x)\simeq -\left[ \frac {\pi \nu }{ \sinh(\pi\nu)}\right]^{\frac 1 2} 
 \frac{\sin[(x-\log (2))\nu+arg\left(\Gamma(1+i\nu)\right)]}\nu; 
\ee 
comparison with the free Schr\"odinger waves gives   
\be 
\int_\R  K_{i\nu}(e^x)K_{i\nu'}(e^x) \, dx= 
\frac{ \pi^2} 
{\sinh(\pi\nu)}\delta\left(\nu^2-{\nu'}^2\right)=N_\lambda^{-2}
\delta(\lambda-\lambda'). 
\ee 
In terms of the original variable $R$ we therefore obtain the 
following normalized generalised eigenfunctions 
\be
\theta_\lambda(R)=N_\lambda R^{\frac {1-d}2}K_{i\nu} (MR).
\label{rmodes}
\ee
The orthonormality and completeness conditions for these modes read
\bea
&&
\int_\X \theta_\lambda(R)  {\theta_{\lambda'}}(R) R^{d-2} {\rm 
d}R = \delta(\lambda-\lambda')
\\
&& \int_{\frac{(d-1)^2}{4} } ^\infty d\lambda\,
\theta_\lambda(R)\theta_\lambda(R') =
R^{-(d-2)}\delta(R-R')
\label{ortho}
\eea 

We now introduce the fields $\Varphi_{\lambda}(y)$ on the de Sitter  
manifold  $\Y$ by smearing the field $\PHI$ with  the complete
set of  radial modes
 (\ref{rmodes}): 
\be 
\Varphi_{\lambda } (y)= \int_{\X} 
\PHI(X) \overline\theta_{\lambda} (R) R^{d-2}dR . 
\label{hu}
\ee  
Our main result is that the field $\Varphi_{\lambda } (y)$ 
is a Klein--Gordon field on the de Sitter manifold in the ``preferred'' 
(also called Euclidean or Bunch-Davies) vacuum state.
In precise terms,  the Minkowski vacuum expectation values (v.e.v.) 
of the fields $\Varphi_{\lambda } (y)$ are given by 
\be
W_{\lambda,\lambda'}(y,y')\equiv 
\langle \Omega|\Varphi_\l(y)\Varphi_{\l'}(y')|\Omega\rangle = \delta(\l-\l') W_\l(y,y'),
\label{restrizione}
\ee
where $W_\l$ is the  ``preferred''  two-point function of de Sitter Klein--Gordon field in dimension $d$ \cite{brosmoschella}. 
In particular,  the fields $\Varphi_\l$ have zero correlation 
(and hence commute) for different values of the square mass $\l$.
 
We now will give an argument to prove the result by first of all
deriving an explicit expression for $W_{\lambda,\lambda'}(y,y')$ (eq. \ref{integral} below) .
Let  us rewrite the v.e.v. appearing at the 
LHS of Eq. (\ref{restrizione}) by using the momentum representation 
of the two-point function of the field $\PHI(X)$:
\bea
&& W_{\lambda,\lambda'}(y,y') = \cr
&& 
= \int_0^\infty \frac {dR}R R^{d-1} \theta_\lambda(R) \int_0^\infty
\frac {dR'}{R'}  {R'}^{d-1} \theta_{\lambda'}(R')  \int \frac{{\rm
d}^{d+1}P}{(2\pi)^{d}}
\delta(P^2-M^2)\Theta(P_0) e^{-iP(X-X')}.
\label{expr}
\eea
In this expression we insert  the  parametrisations 
$X=R\,y$ and $X'=R'\,y'$  and introduce the  vector $\alpha$ defined
by the relation 
$M \alpha = P$, where $P$ is on the mass shell; 
$\alpha$ is therefore  on the unit shell $\alpha^2=1$, 
$\alpha_0>0$.\\
By exchanging the order of integration 
we are led to the following integrals 
(\cite{bateman}, Vol II, Eq. (7.8.5)):
\bea
&& \varphi_\lambda(y,\alpha) = \varphi_\lambda(y\cdot \alpha) = 
M^{\frac {d-1}2} \int_0^\infty e^{i y\cdot\alpha\, MR} \theta_\lambda(R)R^{d-1}
\frac {dR}R = \cr
&& = \sqrt{\frac \pi 2} 
 N_\lambda \Gamma\le(\frac {d-1}2-i\nu
\ri)\Gamma\le(\frac {d-1}2+i\nu\ri)\le((-iy\cdot\alpha)^2-1\ri)^{\frac{2-d}4}
P^{\frac{2-d}2}_{-\frac {1}2 -i\nu}(-iy\cdot\alpha)\ ,\label{plane} 
\eea
where the factor $M^{\frac {d-1}2}$ has been inserted for convenience
so that the function is dimensionless.\\
The functions $\varphi_\lambda(y,\alpha)$  
are a new set of 
{\em plane waves}  on de Sitter manifold i.e. are (global) 
modes satisfying  the
de Sitter Klein--Gordon equation (\ref{kgds}) whose phase 
is constant on planes; $\alpha$  plays the role of wave-vector and 
$P $ is an associated Legendre function \cite{bateman}. 
The integral appearing in the definition of these waves 
is well defined at 
both extrema provided $|\Im(\nu)|< \frac {d-1}2$.\\
We are finally led to consider the following expression:
\be
W_{\lambda,\lambda'}(y,y')=  
\int \frac {{\rm d}^{d+1}\alpha}{(2\pi)^{d}} \delta(\alpha^2-1)\Theta(\alpha_0)
{\overline \varphi}_{\lambda} ( y,\alpha ) \varphi_{\lambda'}(y',\alpha )\ .\label{integral}
\ee
This formula coincides 
with the  ``preferred''   two-point function of the de Sitter 
Klein--Gordon field in dimension $d$ (see Eq. \ref{restrizione}), 
giving at  the same time a new integral 
representation for it. 
The actual full proof of this claim is somewhat involved   
and will be given elsewhere \cite{bertola2}. 

We can however illustrate the result 
in the simplest case $d=1$ where things are easier.
In this case  the de Sitter spacetime 
has only ``time'' and no ``space'': 
it can be visualised as the two branches
of an hyperbola in the two--dimensional Minkowski spacetime; these 
represent the world--lines of  two uniformly accelerated observers.\\
The plane  waves  $\varphi_\lambda$ reduce here  
to ordinary trigonometric functions in the
rectifying parameter of the hyperbola.
  
To carry out the computation 
of the integral (\ref{integral}) 
we  parametrise  $\alpha=
\pmatrix{\cosh(s)\cr 
\sinh(s)} $ and $y(t)=\pmatrix{\sinh(t)\cr \pm \cosh(t)}$;
$t$ is the proper time of
the `freely falling' observer in the one--dimensional de Sitter
spacetime (the same observer is 
actually subject to a constant acceleration when regarded from 
the
Minkowskian ambient space).
By taking the two points  on
the same branch of the hyperbola (say, the right one) we promptly find 
$\alpha\cdot y(t)= \sinh(s+t)$.
The measure ${\rm d}^2 \alpha\delta(\alpha^2-1)\Theta(\alpha_0)$
becomes simply $\frac 1 2 {\rm d} s$. \\
A second step exploits 
the analyticity properties of the Minkowskian Wightman
function and of our plane waves  
by  shifting the 
two times $t,t'$ by an
imaginary part $t \mapsto t-i\pi/2 $  and $t'\mapsto
t'+i\pi/2$.
It follows that 
\beas
&&\int_{\R} \frac{ds}{2(2\pi)}  \int \frac {dR}R\int \frac {dR'}{R'}
 e^{-i\alpha(s)\cdot\le( y(t)-y(t')\ri) }
\theta_\lambda(R)\theta_{\lambda'}(R') \stackrel{\matrix{
t\mapsto t-i\pi/2\cr 
t'\mapsto t'+i\pi/2}}{\longrightarrow} \cr
&&\mapsto
\int_{\R} \frac{ds}{2(2\pi)}  \int \frac {dR}R\int \frac {dR'}{R'}
 e^{- M\le(R \cosh (s+t)+R'\cosh(s+t')\ri) } \theta_\lambda(R)\theta_{\lambda'}(R')
=\cr\vrul
&&=
N_\lambda N_{\lambda'} 
\int_{\R}\frac{ds}{2(2\pi)} \,  \le|\Gamma\le(i\nu\ri)\ri|^2 \cos(\nu (s+t))
\le|\Gamma\le(i\nu'\ri)\ri|^2 \cos(\nu' (s+t'))=\cr\vrul
&&=\frac 14
N_\lambda N_{\lambda'} \le|\Gamma\le(i\nu\ri)\ri|^2
\le|\Gamma\le(i\nu'\ri)\ri|^2  \delta(\nu-\nu')
\cos(\nu(t-t')) =\cr\vrul
&&= \frac 1 {4\nu^2\sinh(\pi \nu)}\delta (\nu-\nu') \cos
(\nu(t-t')) = \delta(\lambda-\lambda') \frac
{\cos(\nu(t-t'))} { 2\nu\sinh(\pi\nu)} 
\eeas
Returning to the Minkowski
spacetime from the correct tubular domains ($t\mapsto t+i\pi/2$
and $t'\mapsto t'-i\pi /2$) we obtain the result
\bes
\langle \Omega |\Varphi_\lambda(y(t))\Varphi_{\lambda'}(y(t')) |
\Omega\rangle = \delta(\lambda-\lambda') \frac
{\cos(\nu(t-t'+i\pi ))} {2 \nu\sinh(\pi\nu)} =
 \delta(\lambda-\lambda') W_\lambda (y(t),y(t'))\ .
\ees
This two--point function correspond to the  ``preferred''  one for 
de Sitter Klein--Gordon quantum field  
as  given 
in \cite{brosmoschella} as well as 
to that of a quantum harmonic oscillator in a thermal state at the
inverse temperature $2\pi$.
Indeed,  
the quantum Klein--Gordon field on a one--dimensional
spacetime corresponds to a single quantum harmonic
oscillator in the Heisenberg picture where the mass represents
the spring constant.
The  thermal time correlation function 
at an inverse temperature $\beta$ of the 
position operator of such as oscillator  is given by:
\be
W(t,t')=\frac {\cos(\omega (t-t' +i\beta/2))}{2\omega \sinh(\omega\beta/2)} 
\label{qo}\ee
which is precisely the expression derived above with $\beta = 2\pi$.
This simple computation gives a quantitative support of the relations
between Hawking effect in de Sitter and Unruh effect in a flat
spacetime as in \cite{bertola,deser}. \\
We can also invert the transformation (\ref{hu}).
Indeed, using the completeness of the radial modes in
eqs. (\ref{ortho}) we can express the ambient field $\PHI$ as 
\be
\PHI(R,y) =\int_{\frac{(d-1)^2}4}^\infty \!\!\!\!\!\! d\lambda\, 
 \theta_\lambda(R) \Varphi_\lambda(y)\ ,
\label{16}
\ee
and consequently obtain the following decomposition of the Wightman
function
\bea
&&\langle\Omega|\PHI(R,y)\PHI(R',y')|\Omega\rangle = \int_{\frac {(d-1)^2}4}^\infty\! \!\!\! d\lambda 
 \int_{\frac {(d-1)^2}4}^\infty \! \!\!\! d\lambda' \,\,
W_{\lambda,\lambda'}(y,y') \theta_{\lambda}(R)
 \theta_{\lambda'}(R')=\cr
&& = \int_{\frac {(d-1)^2}4}^\infty \!\!\!\!\! d\l
\theta_\l(R)\theta_\l(R') W _\l (y,y')\ .
\label{17}
\eea
This formula allows to express the {\em restriction} of the
ambient field $\PHI$ to a fixed leaf $R=R'$ as  a superposition of
Klein--Gordon fields in the respective Euclidean vacuum, 
\bes
\langle\Omega|\PHI(R,y)\PHI(R,y')|\Omega\rangle = \int_{\frac {(d-1)^2}4}^\infty \!\!\!\!\! d\l
|\theta_\l(R)|^2 W^{(E)}_\l (y,y')\ .
\ees
The so--obtained K\"allen--Lehmann type of expansion has  a weight in
the square mass parameter $\l$ given by the density of states per unit
spectrum per unit volume of the self adjoint operator 
$-R^2\le(-\pa^2_R-\frac d R \pa_R + M^2 \ri)$.

These results hold true in any dimension and
can actually be generalised to other manifolds.
For instance, by the same methods, one can decompose a $d+1$-dimensional 
Anti de Sitter (AdS) spacetime into a $d$-dimensional Minkowski spacetime  and 
a $1$-dimensional Minkowski leaf parametrised by a radial coordinate $R$.
In a sense this generalises 
the geometrical basis of  the  AdS/CFT correspondence conjecture 
 \cite{maldacena} where 
one describes the Minkowski space as a boundary of an AdS manifold 
(that is $R$ is sent to infinity).  
These and other results will be discussed elsewhere \cite{bertola2,Bertola:1999sd}.

\end{document}